# High transport $J_c$ in magnetic fields up to 28 T of stainless steel/Ag double sheathed Ba122 tapes fabricated by scalable rolling process

Zhaoshun Gao, Kazumasa Togano, Akiyoshi Matsumoto, and Hiroaki Kumakura*

National Institute for Materials Science, Tsukuba, Ibaraki 305-0047, Japan

**Abstract:**

The recently discovered iron-based superconductors with very high upper critical field ($H_{c2}$) and small anisotropy have been regarded as a potential candidate material for high field applications. However, enhancements of superconducting properties are still needed to boost the successful use of iron-based superconductors in such applications. Here, we propose a new sheath architecture of stainless steel (SS)/Ag double sheath and investigate its influence on the microstructures and $J_c$-$H$ property. We found that the transport $J_c$-$H$ curves for rolled and pressed tapes both show extremely small magnetic field dependence and exceed $3 \times 10^4 \text{A/cm}^2$ under 28 T, which are much higher than those of low-temperature superconductors. More interestingly, 12 cm long rolled tape shows very high homogeneity and sustains $J_c$ as high as $7.7 \times 10^4$ A/cm$^2$ at 10 T. These are the highest values reported so far for iron-based superconducting wires fabricated by scalable rolling process. The microstructure investigations indicate that such high $J_c$ was achieved by higher density of the core and uniform deformation resulting better texturing. These results indicate that our process is very promising for fabricating long Ba122 wires for high field magnet, i.e. above 20 T.

* Author to whom correspondence should be addressed; E-mail: KUMAKURA.Hiroaki@nims.go.jp

The recently discovered iron-based superconductors have attracted great interest as a subject of both fundamental study and practical applications [1-9]. Among the various iron based superconductors, 122-type (AE)Fe$_2$As$_2$ (AE = Sr, Ba) is very attractive for high field applications due to its excellent properties such as very high upper critical field ($H_{c2}$), relatively small anisotropy [10, 11], and an advantageous grain-boundary nature [12]. In order to achieve high critical current density ($J_c$) various processing techniques have been used [13–18]. Among these proposed processes, the powder-in-tube (PIT) method seems the most promising technique for practical applications [19-21]. Thus, the Ba(Sr)122 PIT wires have been investigated intensively [22-31], and high $J_c$ exceeding the level of $10^4$ A/cm$^2$ at 4.2 K and 10 T has been reported for short samples prepared by applying uniaxial pressing [32-35]. However, unfortunately, pressing processes are usually applicable only for a short sample and it is difficult to attain such high $J_c$ values with good reproducibility. For practical applications, it is desirable if high $J_c$ values are attained with good reproducibility by rolling, which is easy to scale up to the long length wire production.

Although silver looks the most suitable sheath material for 122 phase, there are a number of drawbacks of using pure Ag as the sheathing material: poor mechanical properties after the heat treatment that make the coil winding difficult and limit its ability to support Lorentz forces in high magnetic fields, and high occurrence of sausaging during mechanical deformation resulting in inhomogeneous transport current along the conductor length. The sausaging occurs during the cold-working as a result of large differences of hardness between the filament cores and the sheath [36]. In addition to the reduction in effective superconductor cross-sectional area, the resulting roughness of the superconductor/sheath interface reduces the $c$-axis alignment of the Ba122 grain, degrading the overall texture and thus the transport $J_c$.

In this paper, we present preliminary results obtained on Ba122 tapes with a new sheath architecture. The composite sheath consists of a stainless steel (SS) external layer and a protective Ag layer situated between the superconducting ceramic core and the SS layer. The use of a double sheath is effective for improving the mechanical strength of the tapes. In addition, the use of double sheath instead of soft pure Ag suppresses the occurrence of sausaging and improves the grain texture. Furthermore, the configuration of outer SS with high strength and Young's modulus surrounding soft Ag inner sheath keeps the superconducting core in larger compressive stress state during cold rolling and heat treatment processes. This effectively densifies the superconducting core and, thus, enhances $J_c$.

The precursors of $Ba_{0.6}K_{0.4}Fe_2As_{2.1}$ were prepared from Ba filings, K pieces, Fe powder and As pieces [33]. In order to compensate for loss of elements, the starting mixture contained 10-20% excess K. These elemental materials were mixed in an Ar atmosphere about 10 h using a ball-milling machine and placed into a Nb tube of 6 mm outer diameter and 5 mm inner diameter. The Nb tube was put into a stainless steel tube, both ends of which were pressed and sealed by arc welding in an Ar atmosphere for the heat treatment at 900 ºC for 30 h. After heat treatment, the precursor was ground into powder using an agate mortar in a glove box filled with high purity argon gas. The powder was packed into an Ag tube (outside diameter: 8 mm, inside diameter: 3.5 mm), which was subsequently groove rolled into a wire with a rectangular cross section. Then it was drawn into a round wire with the diameter of ~0.9 mm. The wires were deformed into a tape form 0.50 ~ 0.35 mm in thickness using a flat rolling machine. An empty SS tube was flat rolled to prepare a rectangular cross section-tube with a length of 8 cm. The Ag-clad wire was stacked inside this tube which was then rolled to about 0.95 mm in thickness. For comparison, pressed tapes were prepared by cold uniaxial pressing the rolled tapes under a pressure of 2~4 GPa. The rolled and pressed tapes were subjected to a final sintering heat treatment at 850 ºC for 2~4 h. All heat treatments were carried out by putting the samples into a stainless steel tube, both ends of which were pressed and sealed by arc welding in an Ar atmosphere. Table I shows the details of heat treatment for Ba122 tapes. The onset $T_c$ values decrease slightly from ~38.0 K for precursor to ~37.2 K for tapes. The transport critical current $I_c$ at 4.2 K and its magnetic field dependence were evaluated by the standard four-probe method, with a criterion of 1 μV/cm. The transport critical current density, $J_c$, was calculated by dividing $I_c$ by the cross sectional area of the Ba122 core. Magnetic fields up to 12 T were applied parallel to the tape surface. We also carried out $I_c$ measurement in a 28 T hybrid magnet of the Tsukuba Magnet Laboratory (TML) of National Institute for Materials Science. Vickers hardness was measured on the polished cross sections of the tape samples with 0.05 kg load and 10 s duration. We carried out mechanical polishing on the wide surface of the tapes using emery paper and lapping paper, and then Ar ion polishing by cross section polisher to observe microstructures of tapes. After polishing the surface of the samples, we performed scanning electron microscopy (SEM) observations using a SU-70 (Hitachi High-Technologies Corporation).

Figure 1a and b show optical micrographs of the transverse cross sections observed for as-rolled and pressed tapes with SS/Ag double sheaths. After pressing, the tape thickness was reduced from ~0.95 mm to ~0.80 mm. The superconducting core area fraction of the tapes is smaller than Ag sheathed tapes [33]. We can improve the superconducting core area fraction by reducing the thickness of SS tube in the future. A well-defined Ag sheath enclosed in an SS layer can be obviously seen. There is good contact between the Ag and SS. The photograph of rolled tape demonstrates that the composite was deformed uniformly. However, the shape of the core in the rolled tape in Fig 1a is somewhat different from the pressed tape in Fig 1b, with a reduced thickness along the centerline of the pressed tape compared to the rolled one. The uniformity of the core along the length is also essential for the achievement of high transport $J_c$. Optical microstructures of the longitudinal cross sections of the rolled tapes with different sheath materials are compared in Fig 1c and d. The SS/Ag double sheath evidently improves the uniformity of superconducting core thickness with straight core/sheath interfaces. The uniform deformation of superconducting core is attributable to the increase in the hardness of sheath materials [36, 37]. Furthermore, the uniform Ag/superconductor interface can improve the degree of local texture of the superconducting grains [38].

The texturing was studied by X-ray diffraction analyses which were performed on the wide surface of the tapes after the sheaths were mechanically removed. Fig. 2 shows the XRD patterns of the randomly orientated precursor and the tapes processed by flat rolling and uniaxial pressing. It is found that all samples consist of the major phase of $Ba_{1-x}K_xFe_2As_2$ except for the Ag and SS peaks from the sheath materials. The strong intensities of the (*00l*) peaks of $Ba_{1-x}K_xFe_2As_2$ phase were observed in tape samples. This clearly demonstrates that a crystalline orientation of the *c*-axis is well developed in Ba122 tapes by cold rolling and pressing procedures. The texture of the samples was evaluated from XRD patterns, using the texture factor F, defined as [39].

$$F=(\rho - \rho_0)/(1 - \rho_0),$$

Where $\rho_0 =\sum I_0(00l)/ \sum I_0(hkl)$, $\rho =\sum I(00l)/ \sum I(hkl)$, $I$ and $I_0$ are the intensities of each reflection peak (*hkl*) for the oriented and random samples, respectively. The c-axis orientation factors F for rolled and pressed samples are 0.71 and 0.85, respectively. These are much higher than the previous reports [28, 33-35]. This result indicates that the use of hard SS as an outer sheath is effective to improve grain alignment of Ba122 tapes.

The densification of superconducting core is crucial for high transport $J_c$. In this work, we used Vickers hardness measurements as an indicator for density of the core and investigated the relationships among hardness, sheath materials, fabrication process, and $J_c$ values. Fig. 3 plots $J_c$ (10 T, 4.2 K) versus microhardness for both rolled and pressed tapes. There is a strong linear relation between the hardness of superconducting core and $J_c$. As the hardness increased, the $J_c$ of the Ba122 tapes also increased. Furthermore, it appears that the SS/Ag sheathed tapes had harder cores than the Ag sheathed tapes for both rolling and pressing, indicating the effectiveness of using a SS outer sheath.

Wire applications of superconductor require large transport critical current density in high magnetic fields. Figure 4 displays that both rolled and pressed Ba-122 tapes with a SS/Ag double sheath show remarkable in-field $J_c$ performance at 4.2 K in comparison to the values for commercial superconductors NbTi, $Nb_3Sn$ [40] and PIT processed $MgB_2$ [41]. The measurement was initially carried out using a 12 T superconducting magnet, and then a 28 T hybrid magnet. The samples were stored for a week between measurements. The data of the hybrid magnet and the 12 T superconducting magnet measurements coincide with each other indicating that no degradation occurred during one week storage and the thermal cycles. Due to an extremely small magnetic field dependence, the critical current density $J_c$ is still of the order of $10^4$ A/cm$^2$ in high magnetic fields beyond 20 T, the best value being $3.2\times10^4$ A/cm$^2$ at 28 T. To our knowledge this is the highest critical current density for iron based superconducting wires ever reported for such high magnetic fields. Quite interestingly, a crossover with Nb$_3$Sn is around 16 T. This clearly reveals the potential of this kind of material for use in very high-field superconducting magnets.

From the viewpoint of applications, the long tapes produced by the scalable, industrial rolling technique is more important than the short tapes prepared by the more academic method of pressing. Two 12 cm long SS/Ag double sheathed tapes were fabricated to check the possibility of producing long length Ba122 wires by our process. Figure 5 shows the variation of critical current densities along the 12 cm length of the tapes, both showing less than 6% fluctuations. Furthermore, a very high $J_c$ of $7.7\times10^4$ A/cm$^2$ was measured at 4.2 K and 10 T. The $J_c$ value obtained here is by far the highest ever reported for flat-rolled iron-based tapes. These results demonstrate that our process is quite stable and reproducible for fabrication of long and very homogeneous tapes with high performance.

In order to investigate the influence of sheath materials on the samples, we studied the difference in microstructures of superconducting cores. Figure 6 shows typical SEM images for the rolled and pressed SS/Ag sheathed tapes. For comparison, the pictures for Ag sheathed tapes are also included in the figure. The typical grain size for all samples is 5 ~ 10 μm. Both SS/Ag double-sheathed samples display a denser microstructure than the rolled tape with Ag sheath. This result is in agreement with the hardness analysis. High strength and Young's modulus of SS result in higher compression stress during cold rolling and heat treatment processes. Thus, the microstructure in the SS/Ag double sheathed samples is denser than that in the pure Ag sheathed tapes. It should be noted that SS/Ag double sheathed samples were subjected to the final heat treatment without intermediate annealing. The fact that the high $J_c$ values can be obtained in a single heat-treatment is rather interesting both from technical and economical points of view. In present study, we only investigate the influence of SS/Ag sheath on sample's density. However the grain shape, size and intergranular structure between neighboring grains also have important influence on the texturing degree and grain connection of the tapes. It is worth to study in the future.

Excellent transport $J_c$ values of $\sim 10^4 A/cm^2$ under magnetic fields up to 28 T were obtained in SS/Ag double sheathed $Ba_{1-x}K_xFe_2As_2$ tapes. We propose that this remarkable high $J_c$ is due to the stiffness of SS. In the case of Ag-clad tapes, Ag is too soft at high temperature to keep the ceramic core in compressive state and allows its expansion. Therefore, additional heat-treatments and intermediate densifications must be performed in order to increase the density of the ceramic core [32]. Recent reports on the use of hot pressing partly solve this problem [34, 35]. The high strength and Young's modulus of SS result in higher compressive stress during heat treatment and prevents the development of porosity. This effectively densifies the superconducting core as illustrated in figure 6 and, thus, enhances $J_c$.

The influence of sheath materials on the cold mechanical process is another key point for this remarkable result. According to the powder-flow model of Han and Freltoft [42], the final powder packing density depends on the properties of the powder and the sheath material. A high yield stress of SS sheath can provide sufficient stress during the cold-working process to increase the powder compaction and consequently improve the grain connectivity. The mechanical deformation process of Ba122 tapes not only changes the powder density, but may also lead to variation in core layer thickness (sausaging). Sausaging is caused by ill-matched workability of the ceramic superconductor and the sheath material. This kind of problem is very serious from the viewpoint of fabrication of long, very homogeneous tapes with high performance. Fig2c and d clearly demonstrate that the sausaging effect was greatly suppressed by using of SS/Ag double sheath compared to soft pure Ag. In addition, the smooth interface is favorable to improve the grain texture and thus the transport $J_c$. This is confirmed in figure 5, which shows high transport $J_c$ values and excellent longitudinal homogeneity of 12 cm long samples.

Although the pressed tapes have higher $J_c$ values than the rolled tapes, the $J_c$ gap between the pressed and rolled tapes is greatly reduced by adopting SS/Ag double sheath. Furthermore, rolling is easy to scale up to long-length wires. Our previous study showed that the higher $J_c$ values of the pressed tapes are mainly due to the higher core density, more aligned grains and a change in the microcrack structure [33]. This indicates that if we can further improve the core density and grain texture, and also reduce microcracks in rolled tapes, higher $J_c$ performance can be achieved in the future.

In conclusion, the transport $J_c$-$H$ curves for rolled and pressed tapes both show extremely small magnetic field dependence and exceed $3\times10^4 \text{A/cm}^2$ under 28 T, which could not be reached by conventional NbTi and Nb$_3$Sn conductors. More interestingly, 12 cm long rolled tape shows very high homogeneity and sustains $J_c$ as high as $7.7\times10^4$ A/cm$^2$ at 10 T. These are the highest values reported so far for iron-based superconducting wires fabricated by rolling process. We believe that this simple and scalable process is very promising for fabricating long length Ba122 wires for high field magnets.

## Acknowledgements

This work was supported by the Japan Society for the Promotion of Science (JSPS) through its "Funding Program for World-Leading Innovative R&D on Science and Technology (FIRST Program)". We acknowledge Dr. H. Fujii, Dr. H. Takeya, Dr. S. J. Ye and Miss Y. C. Zhang of the National Institute for Materials Science for their assistance in the experiments.

# References

[1] Y. Kamihara, T. Watanabe, M. Hirano, H. Hosono, J. Am. Chem. Soc. **130**, 3296 (2008).

[2] X. H. Chen, T. Wu, G. Wu, R. H. Liu, H. Chen, D. F. Fang, Nature **453**, 761 (2008).

[3] H. H. Wen, G. Mu, L. Fang, H. Yang, and X. Zhu, Europhys. Lett. **82**, 17009 (2008).

[4] Z. A. Ren, W. Lu, J. Yang, W. Yi, X. L. Shen, Z. C. Li, G. C. Che, X. L. Dong, L. L. Sun, F. Zhou, Z. X. Zhao, Chin. Phys. Lett. **25**, 2215 (2008).

[5] M. Rotter, M. Tegel, D. Johrendt, Phys. Rev. Lett. **101**, 107006 (2008).

[6] X. Wang, S. R. Ghorbani, G. Peleckis, and S. Dou, Adv. Mater. **21**, 236 (2009).

[7] M. Putti, I. Pallecchi, E. Bellingeri, M. R. Cimberle, M. Tropeano, C. Ferdeghini, A. Palenzona, C. Tarantini, A. Yamamoto, J. Jiang et al., Supercond. Sci. Technol. **23**, 034003 (2010).

[8] J. H. Durrell, C. B. Eom, A. Gurevich, E. E. Hellstrom, C. Tarantini, A. Yamamoto, and D. C. Larbalestier, Rep. Prog. Phys. **74**, 124511 (2011).

[9] J. Shimoyama, Supercond. Sci. Technol. **27**, 044002 (2014).

[10] A. Yamamoto, J. Jaroszynski, C. Tarantini, L. Balicas, J. Jiang, A. Gurevich, D. C. Larbalestier, R. Jin, A. S. Sefat, M. A. McGuire, B. C. Sales, D. K. Christen, and D. Mandrus, Appl. Phys. Lett. **94**, 062511 (2009).

[11] H. Q. Yuan, J. Singleton, F. F. Balakirev, S. A. Baily, G. F. Chen, J. L. Luo, N. L. Wang, Nature **457**, 565 (2009).

[12] T. Katase, Y. Ishimaru, A. Tsukamoto, H. Hiramatsu, T. Kamiya, K. Tanabe, and H. Hosono, Nat. Commun. **2**, 409 (2011).

[13] Z. S. Gao, L. Wang, Y. P. Qi, D. L. Wang, X. P. Zhang, Y. W. Ma, Supercond. Sci. Technol. **21**, 105024 (2008).

[14] Y. Mizuguchi, K. Deguchi, S. Tsuda, T. Yamaguchi, H. Takeya, H. Kumakura, and Y. Takano, Appl. Phys. Express **2**, 083004 (2009).

[15] L. Fang, Y. Jia, V. Mishra, C. Chaparro, V. K. Vlasko-Vlasov, A. E. Koshelev, U. Welp, G. W. Crabtree, S. Zhu, N. D. Zhigadlo, S. Katrych, J. Karpinski, W. K.

Kwok, Nat. Commun. **4**, 2655 (2013).

[16] C. Tarantini, S. Lee, Y. Zhang, J. Jiang, C. W. Bark, J. D. Weiss, A. Polyanskii, C. T. Nelson, H. W. Jang, C. M. Folkman, S. H. Baek, X. Q. Pan, A. Gurevich, E. E. Hellstrom, C. B. Eom, and D. C. Larbalestier, Appl. Phys. Lett. **96**, 142510 (2010).

[17] W. Si, S. Jung Han, X. Shi, S. N. Ehrlich, J. Jaroszynski, A. Goyal, and Q. Li, Nat. Commun. **4**, 2337 (2013).

[18] S. Trommler, J. Hänisch, V. Matias, R. Hühne, E. Reich, K. Iida, S. Haindl, L. Schultz, and B. Holzapfel, Supercond. Sci. Technol. **25**, 084019 (2012).

[19] Z. S. Gao, L. Wang, Y. P. Qi, D. L. Wang, X. P. Zhang, Y. W. Ma, H. Yang, and H. H. Wen, Supercond. Sci. Technol. **21**, 112001 (2008).

[20] Y. P. Qi, X. P. Zhang, Z. S. Gao, Z. Y. Zhang, L. Wang, D. L. Wang, and Y. W. Ma, Physica C **469**, 717 (2009).

[21] Y. W. Ma, Z. S. Gao, Y. P. Qi, X. P. Zhang, L. Wang, Z. Y. Zhang, D. L. Wang, Physica C **469**, 651 (2009).

[22] K. Togano, A. Matsumoto, and H. Kumakura, Appl. Phys. Express **4**, 043101 (2011).

[23] L. Wang, Y. P. Qi, X. P. Zhang, D. L. Wang, Z. S. Gao, C. L. Wang, C. Yao, and Y. W. Ma, Physica C **471**, 1689 (2011).

[24] Z. Gao, L. Wang, C. Yao, Y. Qi, C. Wang, X. Zhang, D. Wang, C. Wang, and Y. Ma, Appl. Phys. Lett. **99**, 242506 (2011).

[25] K. Togano, A. Matsumoto, and H. Kumakura, Solid State Commun. **152**, 740 (2012).

[26] J. D. Weiss, C. Tarantini, J. Jiang, F. Kametani, A. A. Polyanskii, D. C. Larbalestier, and E. E. Hellstrom, Nature Mater. **11**, 682 (2012).

[27] Z. Gao, Y. Ma, C. Yao, X. Zhang, C. Wang, D. Wang, S. Awaji, and K. Watanabe, Sci. Rep. **2**, 998 (2012).

[28] Y. Ma, Supercond. Sci. Technol. **25**, 113001 (2012).

[29] A. Matsumoto, Z. Gao, K. Togano, and H. Kumakura, Supercond. Sci. Technol. **27**, 025011 (2014).

[30] S. Pyon, Y. Tsuchiya, H. Inoue, H. Kajitani, N. Koizumi, S. Awaji, K.

Watanabe and T. Tamegai, Supercond. Sci. Technol. **27**, 095002 (2014).

[31] K. Togano, Z. Gao, A. Matsumoto, and H. Kumakura, Supercond. Sci. Technol. **26**, 065003 (2013).

[32] K. Togano, Z. Gao, A. Matsumoto, and H. Kumakura, Supercond. Sci. Technol. **26**, 115007 (2013).

[33] Z. Gao, K. Togano, A. Matsumoto, and H. Kumakura, Sci. Rep. **4**, 4065 (2014).

[34] H. Lin, C. Yao, X. Zhang, H. Zhang, D. Wang, Q. Zhang, Y. Ma, S. Awaji, and K. Watanabe, Sci. Rep. **4**, 4465 (2014).

[35] X. Zhang, C. Yao, H. Lin, Y. Cai, Z. Chen, J. Li, C. Dong, Q. Zhang, D. Wang, Y. Ma, H. Oguro, S. Awaji, and K. Watanabe, Appl. Phys. Lett. **104**, 202601 (2014).

[36] Z. Han, P. Skov-Hansen, and T. Freltoft, Supercond. Sci. Technol. **10**, 371 (1997).

[37] K. Osamura, S. S. Oh and S. Ochiai, Supercond. Sci. Technol. **5**, 1 (1992).

[38] M. Dao, R. Asaro, R. Sebring, Phil. Mag. A **78**, 857 (1995).

[39] F. Lotgering, J. Inorg. Nucl. Chem. **9**, 113 (1959).

[40] P. J. Lee, http://www.magnet.fsu.edu/magnettechnology/research/asc/plots.html

[41] H. Kumakura, H. Kitaguchi, A. Matsumoto, H. Hatakeyama, IEEE Trans. Appl. Supercon. **15**, 3184 (2005).

[42] Z. Han, and T. Freltoft, Appl. Supercond. **2**, 201 (1994).

TABLE I. The parameters of Ba122 tapes.

| Sheath | Cold work | Hardness (average) | Intermediate heat-treatment | Final heat-treatment |
|---|---|---|---|---|
| SS/Ag | Pressed | 193.4 | No | 850 ºC×2h |
| | Rolled | 156.7 | No | 850 ºC×2h |
| Ag | Pressed | 150.7 | 800 ºC×2h | 850 ºC×4h |
| | Rolled | 94.9 | 800 ºC×2h | 850 ºC×4h |

# Figure captions

Figure 1  Optical micrographs of the transverse cross sections observed for (a) rolled and (b) pressed tapes with SS/Ag double sheaths. The longitudinal cross sections of the rolled tapes with (c) SS/Ag double sheath and (d) Ag sheath.

Figure 2  X-ray diffraction patterns for $Ba_{1-x}K_xFe_2As_2$ random powder, rolled and pressed tapes with SS/Ag double sheath.

Figure 3  The $J_c$ (10 T, 4.2 K) as a function of hardness for rolled and pressed tapes.

Figure 4  Transport $J_c$ values plotted as a function of applied magnetic field. $J_c$ of Ba122 tapes are compared to commercial NbTi, $Nb_3Sn$ [40] and $MgB_2$ [41] wires.

Figure 5  $J_c$ homogeneity of 12 cm long rolled tapes measured at 4.2 K and 10 T.

Figure 6  SEM images of pressed tape with (a) SS/Ag double sheath and (b) Ag sheath and rolled tapes with (c) SS/Ag double sheath and (d) Ag sheath.

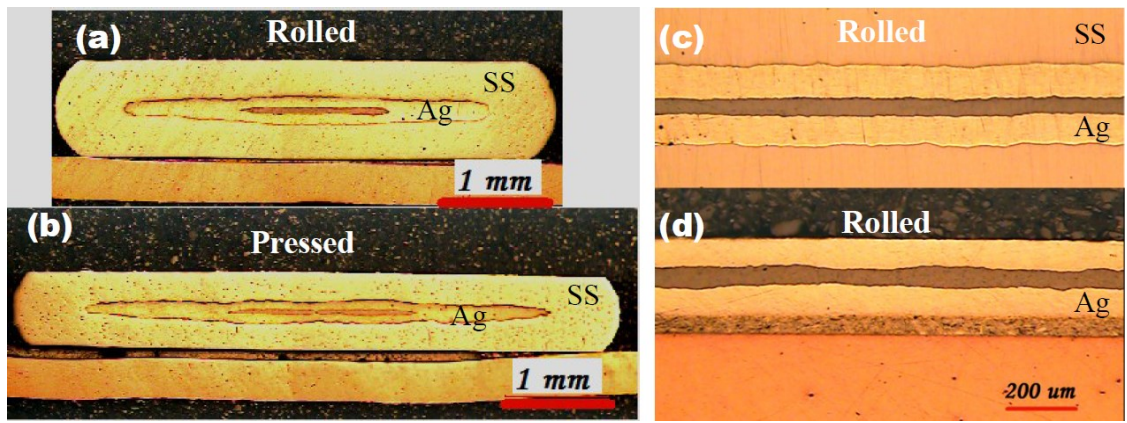

Fig.1 Gao et al.

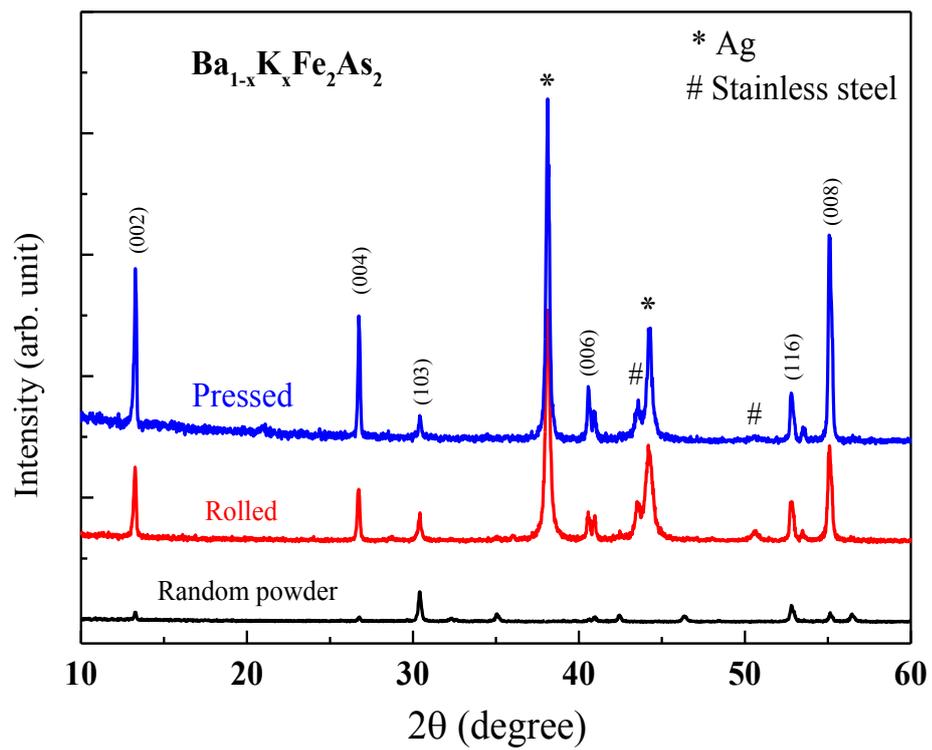

Fig.2 Gao et al.

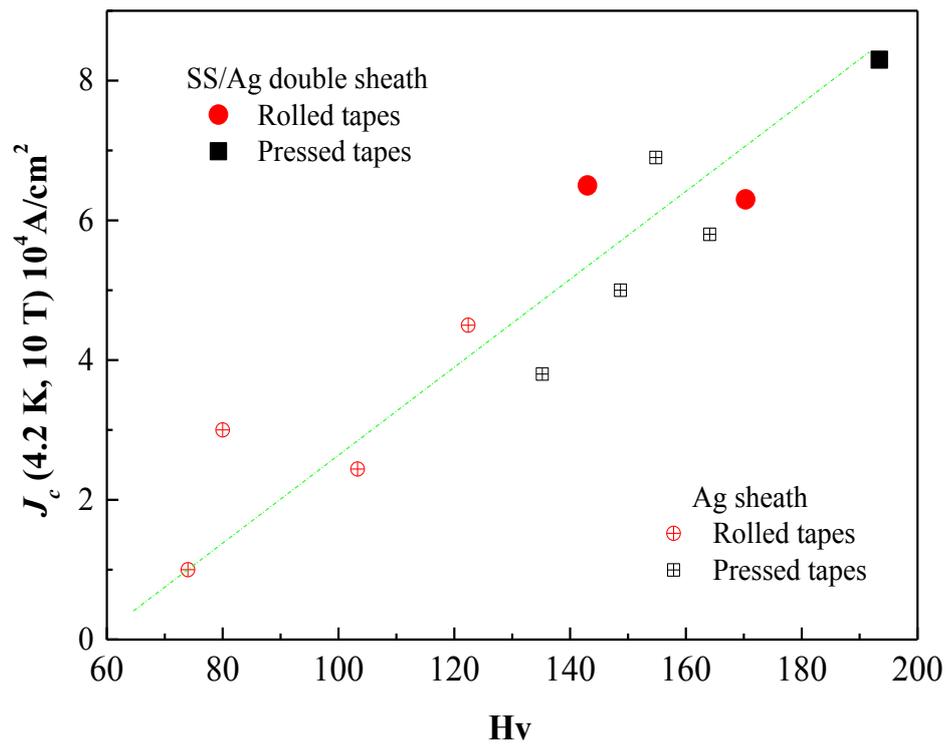

Fig.3 Gao et al.

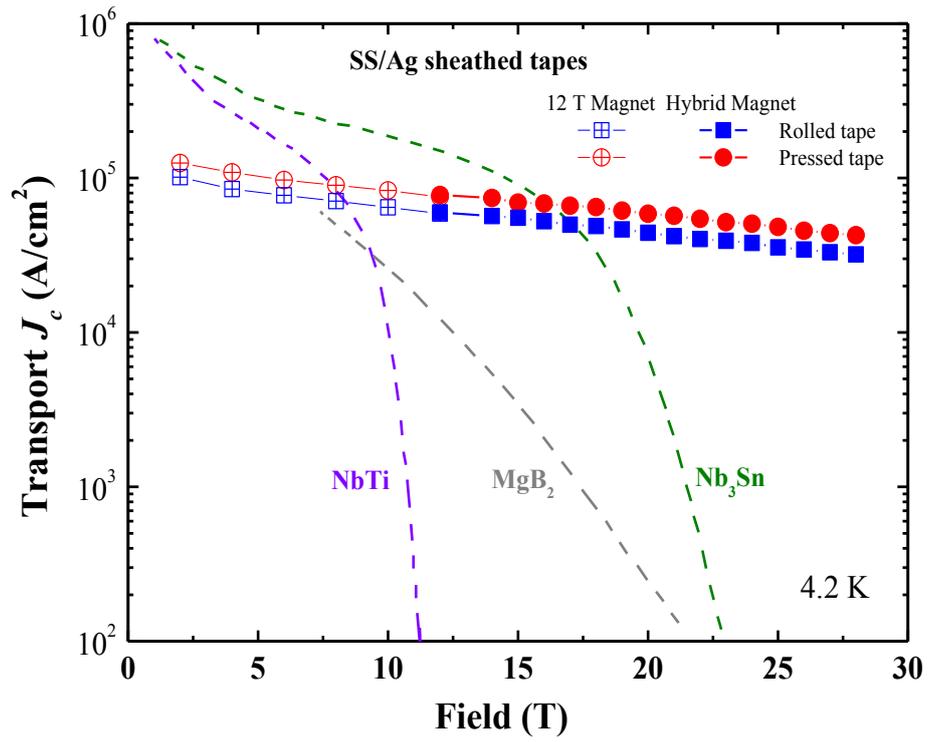

Fig.4 Gao et al.

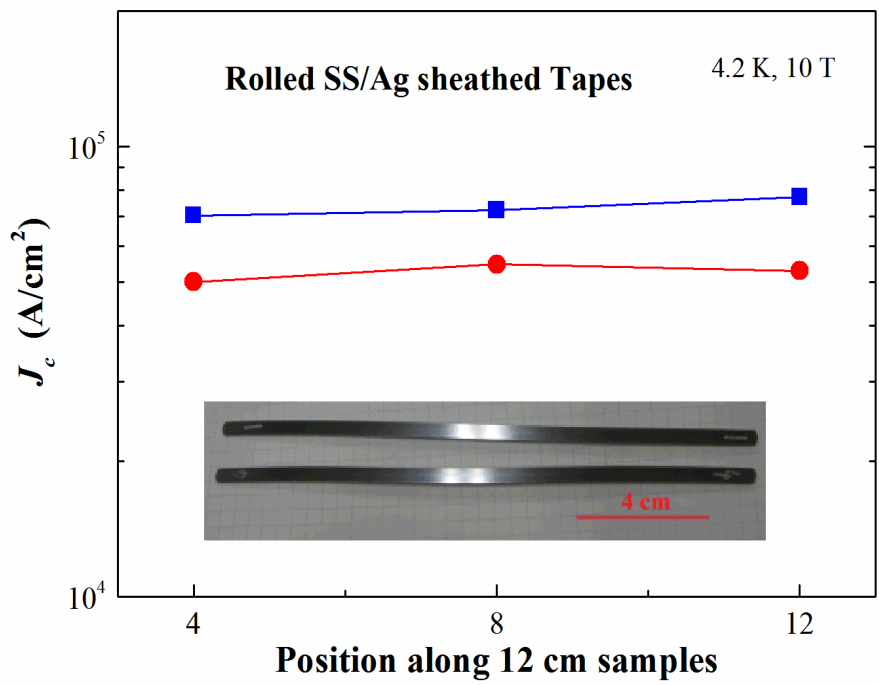

Fig.5 Gao et al.

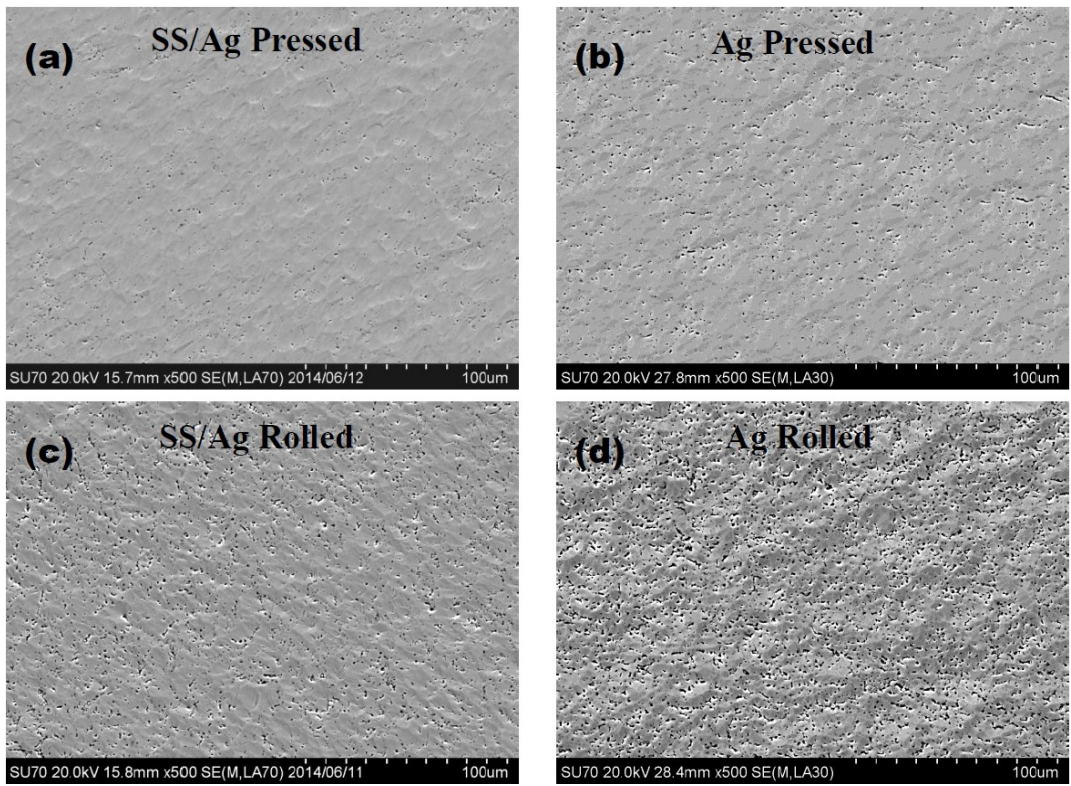

Fig.6 Gao et al.